\documentstyle[preprint,eqsecnum,aps]{revtex}
\tightenlines

\catcode`\@=11

\catcode`\@=11

\begin{document}

\begin{flushright} 

IFUP-TH 34/97
 \end{flushright}
\vskip 25pt plus 3pt minus 3pt

\begin{center}
  {\bf DYNAMICAL EFFECTS ON THE PARALLEL MOMENTUM DISTRIBUTIONS OF NEUTRONS
  FROM  HALO  BREAKUP} 
\end{center}
\vspace{.6em}
\begin{center}
{Angela Bonaccorso$^{+}$ and David M.Brink$^{\dag}$}

{\footnotesize {\it $^+$Istituto Nazionale di Fisica Nucleare,
 Sezione di Pisa, 56100 Pisa, Italy,}

{\small E-mail:  BONACCORSO@AXPIA.PI.INFN.IT}

{\it $^{\dag}$ Dipartimento di Fisica, Universit\`a di Trento, 38050
Povo, Italy,
 and

ECT$^*$, 38050 Villazzano, Trento, Italy,} {\small 
E-mail: BRINK@SCIENCE.UNITN.IT}}

\end{center}
{\footnotesize 
\begin{center}
{\bf Abstract}
\end {center}
In this paper we study the energy spectra and related parallel
momentum distributions of the neutrons from the breakup of
$^{11}Be$. 
Earlier papers on   transfer to the continuum reactions have shown
that the breakup amplitude reduces to a simple eikonal approximation in
the limit of weakly bound projectiles. The aim is to establish a
reliable method to obtain information on the structure of the halo
from the experimental results which depend on the reaction mechanism. 
  Theoretical results for the widths of the parallel momentum distributions and
for the total breakup cross section are compared to recent experimental data.

\begin{flushleft} {\bf PACS }
number(s):25.70.Hi, 21.10Gv,25.60Ge,25.70Mn,27.20+n
 \end{flushleft}}

\newpage

\section
{\bf Introduction}\label{int}

The fact that energy spectra of particles emitted to the continuum
from a nucleus in a nuclear reaction bear some relation with their
original momentum distribution has been known and studied for many
years \cite{ser}-\cite{sl} and the form and widths of the energy
distribution have been related to the original momentum distribution
of the emitted particle in the projectile.  Reactions with final state
having a neutron in the continuum have been studied in detail by us
\cite{bb}-\cite{ab} and we arrived at the conclusion that
there were dynamical effects modifying the original
distributions\cite{tiina}.
 An important consequence of our approach is that it is the
component of the momentum distribution of the breakup neutron in the
projectile parallel to the direction of relative motion that is
responsible for the form of the energy distribution. This is in
contrast to the results given in refs.\cite{ser,gla,esb}  where it is the total
momentum distribution of the neutron in the target which is important while it
is similar to the approaches of \cite{esb1,hans,ot}.
 Ideas and techniques very similar to ours have been recently used in
\cite{hans} to the study of halo breakup.

	We will not repeat derivations given in earlier papers but, in
order to make this paper more self contained, we will describe the
physical motivation of our approach and the underlying assumptions.
Our aim is to show that halo breakup can be studied by using simple analytical
formulae which are more general and more accurate than the eikonal forms and
reduce to them in the limit of zero initial binding energy. The main difference
between the present approach and the eikonal approach comes from the fact
that the rescattering of the breakup neutron on the target is calculated by the
optical model. On the other hand the use of the semiclassical
approximation for the relative motion of the two ions, discussed in the
following, makes our method computationally much simpler than the DWBA type of
approach used for example in Ref.\cite{sl}.

 One basic assumption is the masses of the target and
projectile core are large compared with that of the neutron. Another is that the
relative motion of the projectile and target can be treated classically. For the
high energy reactions described in the present paper we assume that the center
of the projectile core moves on a straight line path with constant velocity and
with a particular impact parameter relative to the target. In our reference
frame the target is at rest. We represent the interaction of the neutron with the
projectile core by a single particle potential moving with the projectile
velocity, and its interaction with the target by an optical potential.  We use
an approximate solution of the time dependent Schr\"odinger equation for the
neutron moving under the influence of these two potentials with the initial
condition that it is bound in a particular state in the projectile.  The theory
presented in Section II yields an expression Eq.(\ref{dpde}) for the momentum
(or energy) distribution of the neutron in the final state as a function of the
impact parameter. The cross-section is obtained by integrating over the impact
parameter.
	
	One consequence of the use of a straight line trajectory with
constant velocity for the motion of the projectile core relative to
the target is that the theory does not satisfy the overall energy and center of
mass momentum conservation conditions. The neutron receives energy from the time
dependent potential fields associated with the relative motion of the target and
projectile. For a given neutron final energy and momentum the overall energy and
momentum conditions have to be put in by hand.  We discuss this point at
the beginning of the next section. The use of straight line trajectories
neglects Coulomb deflections. This is a good approximation for light targets
and projectiles and high incident energies. We have discussed
corrections for Coulomb deflections in earlier papers \cite{27} on transfer to
bound states. The same methods could be used for breakup where
necessary.

\section
{\bf The breakup cross section}\label{two}

	The breakup theory presented in this paper evolved from a
theory for neutron transfer to resonance states of the target \cite{bb}. In
that case the final neutron energy $\varepsilon_f$ was the energy of
the neutron resonance in the target. Thus in the present work
$\varepsilon_f$ is the energy of the neutron relative to the target in
the final state.  For elastic breakup this is the same as the final
laboratory energy of the neutron if the target recoil kinetic energy is
neglected. In the case of compound nucleus formation $\varepsilon_f$ is the
excitation energy of the compound state above the neutron threshold in
the residual nucleus. In the case of inelastic scattering it is the
energy of the breakup neutron before it scatters from the target. This
is equivalent to the sum of the excitation energy of the target final state
and the final neutron energy relative to the target. If
the target recoil kinetic energy is neglected the final kinetic energy $E_f$ of
the ejectile is given by the energy conservation condition
 \begin{equation} E_f-E_{inc}= Q 
=\varepsilon_i-\varepsilon_f ,\label{ec}\end{equation}
 where $E_{inc}$ is the initial incident energy of the projectile in
the laboratory, Q is the reaction Q-value given by
$Q=\varepsilon_i-\varepsilon_f$ and $\varepsilon_i$ is the initial neutron
binding energy in the projectile.  The projectile is supposed to remain in its
ground state after the transfer.  With this approximation Eq.(\ref{ec}) relates
$\varepsilon_f$ to the energy loss spectrum of the ejectile. It is not possible
to calculate corrections due to target recoil within the present theory because
of the three-body character of the final state. This should be done to discuss
experiments in which a full kinematical reconstruction is made. However such
experiments are not available at present.

	The starting point of the present paper is the expression for
the energy distribution of the breakup neutron given in Eq.(\ref{dpde}) 
which was derived in \cite{bb}.  It is obtained by working in the
target reference frame.  The projectile-target relative motion is
treated semiclassically by using a trajectory of the center of the
projectile relative to the center of the target ${\bf s}(t)={\bf
d}+{\bf v}t$ with constant velocity v in the z-direction and impact
parameter {\bf d} in the xy-plane.

Eq.(\ref{dpde}) gives the neutron transfer probability from a definite single
particle state of energy
 $\varepsilon_i$, momentum $\gamma_i=\sqrt
{-2m\varepsilon_i}/\hbar$, and angular momentum $l_i$  in the
projectile to a final continuum state of energy $\varepsilon_f$,
momentum $k_f=\sqrt {2m\varepsilon_f}/\hbar$ within an interval
$d\varepsilon_f$. It is the sum of the transfer probabilities to
each possible final $l_f$-state in  the energy bin $d\varepsilon_f$

\begin{equation}{dP\over d\varepsilon_f}
\approx \Sigma_{l_f}(|1-\langle S_{l_f}\rangle |^2+1-|\langle
S_{l_f}\rangle |^2) B(l_f,l_i).  \label{dpde}\end{equation}

The physical interpretation is that the
projectile brings up the neutron which is scattered into a continuum
state by the target.  The interaction of the neutron with the target is
represented by the S-matrix for scattering of a free neutron by the
target nucleus.

In the derivation of Eq.(\ref{dpde}) in \cite{bb} the neutron final state was
taken as a scattering wave function of energy $\varepsilon_f$, including all
possible neutron-target final state interactions,
 and
$\langle S_{l_f}\rangle $ is the optical model, energy averaged
S-matrix which describes the re-scattering of the neutron on the
target\cite {bb3}.  In Ref.\cite {bb1}  it was shown that the first term of
 Eq.(\ref{dpde}), proportional to
 $|1-\langle S_{l_f}\rangle |^2$ , gives the neutron elastic breakup
spectrum while the second term proportional to the transmission
coefficient $T=1-|\langle S_{l_f}\rangle |^2$ gives the absorption
spectrum. 
 This term contains contributions from inelastic scattering of the
breakup neutron by the target nucleus and also from compound nucleus
formation.

The factor $B(l_f,l_i)$ is an elementary transfer
probability which depends on the details of the initial and final
states, on the energy of relative motion and on the distance of
closest approach $d$ between the two nuclei. Its explicit expression
reads:

\begin{equation} B(l_f,l_i)={1\over 2}\left({\hbar\over mv}
\right)^2{m\over \hbar^2 k_f}(2l_f+1)  P_{l_f}(X_f) \vert C_1\vert
^2{e^{-2\eta d} \over 2\eta d}  P_{l_i}(X_i),\label{B}
 \end{equation}
 where   the arguments of the Legendre polynomials $P_{l_i}$ and
$P_{l_f}$ are respectively
$X_i=1+2(k_1/\gamma_i)^2$ and $X_f=2(k_2/k_f)^2-1$. Also
$k_1=-(\varepsilon_i-\varepsilon_f+{1\over 2}mv^2)/(\hbar v)$ and
$k_2=-(\varepsilon_i-\varepsilon_f-{1\over 2}mv^2)/(\hbar v)$ are the
 z-components of the neutron momentum in the initial and final state
respectively. $\eta$ is the modulus of the transverse component of the
neutron momentum.  It is conserved during the breakup process
and in fact it can be given both in terms of the initial or final
neutron parameters as $\eta^2=k_1^2+\gamma_i^2=k_2^2-k_f^2$. $mv^2/2$ is the incident energy per nucleon at the
distance of closest approach $d$ for the ion-ion collision.

In the case of a weakly bound projectile it is possible to obtain much
simpler expressions for the elastic breakup and absorption
probabilities. In fact $k_2$ which is the $k_z$ component of the
neutron momentum with respect to target, in the limit of a very small
initial binding energy takes the value  $k_2\approx k_f$ . Then the 
Legendre polynomial $P_{l_f}\approx 1$ since its argument is very
close to one. By introducing the classical angular momentum 
$\lambda=k_fb$ and $2l_f+1=2(l_f+1/2)=2\lambda$ in  Eq.(\ref{dpde})
we can replace the sum over the final angular momenta with an integral
over the neutron impact parameters $b$ with respect to the target.
Also when the neutron
rescattering on the target  takes place at relatively high energy
($\varepsilon_f\simeq {1\over 2}mv^2$),  the phase shift in 
$S_{l_f}=e^{2i\delta_{l_f}}$ can
be approximated by the eikonal form using a
phenomenological optical potential such that
$ S_{l_f}\approx e^{{-i\chi({ b})}}$ and
$\chi({ b})={1\over \hbar v}\int_{-\infty}^{\infty} 
V_2(x,y,z^{\prime})dz^{\prime}$
where $V_2$ is a complex potential whose real and imaginary
strengths are negative. Then
 \begin{eqnarray}{dP(d)\over d \varepsilon_f}&\approx &
{m\over \hbar^2 k_f}\int_0^\infty bdb
\left[|(1-e^{-i\chi({ b})})|^2 + 1-|e^{-i\chi({ b})}|^2\right]
 |\tilde {\psi}_1
 ( d ,k_1)|^2 ,\label{pg}\end{eqnarray}
which is the product of
the free neutron elastic (plus absorption) cross section, by the modulus
square of  the initial state momentum distribution along the
relative motion direction,  for a fixed value of the distance $d$. 
The more accurate  expression used in \cite{ab} corresponds to 
replacing the $d$ in the wave function $\tilde
{\psi}_1$ in the last term of Eq.(\ref{pg}) by $|\bf {b - d}|$. It
can be obtained by approximating  
$P_{l_f}$ by  a Bessel function, $P_{l_f}(X_f)\approx
I_0(2\eta l_f/k_f)= I_0(2\eta b)$. 
 The total breakup probability is the sum of an elastic
plus absorptive term because the  full final state  wave function
contains a
 S-matrix which is unitary. In   \cite{ab} instead we started directly
from an eikonal wave function for the neutron final state and we
obtained only the elastic breakup term since in that case the S-matrix
was not unitary. The results discussed in this paper were obtained with
Eq.(\ref{dpde}), but Eq.(\ref{pg}) is very useful to get an insight into the
physics involved in halo breakup.

In Eq.(\ref{pg}) $|\tilde {\psi}_1|^2 $ is the initial state momentum
distribution \cite{ab}
\begin{equation}|\tilde {\psi}_1
 ( d ,k_1)|^2 ={1\over 2l_i+1}\Sigma_{m_i}|\tilde \psi_{l_im_i}(d ,k_1)|^2
\approx |C_1|^2~ {e^{-2\eta d}\over 2\eta d}P_{l_i}(X_i)
\label{psi} ,\end{equation}
 $C_1$ is the asymptotic normalization constant of the initial
wave function.
 The energy
distribution of the final neutron has its main origin in the  $ k_1$
dependence of  $\tilde {\psi}_1$.

 The relation between
the energy distribution probability and cross section is given by
 \begin{equation}
 {d\sigma\over d \varepsilon_f}=C^2S
 \int_0^{\infty} d^2{\bf d} {d P(d) \over d \varepsilon_f} 
 P_{el}(d) \label{cross} \end{equation}
 as in  Eq.(7) of \cite{ab}. 
$P(\varepsilon_f,d)$ is given by Eq.(\ref{dpde}).
$P_{el}(d)=|S_{ct}|^2$ is the ion-ion elastic scattering probability given by
the modulus square of the projectile core-target S-matrix. The above factorized
form of the neutron-target  scattering by the core-target scattering has been
used already in several papers on halo breakup , as Ref.\cite{esb1,ot,be,ab} and
 and it is well known to hold for peripheral reactions\cite{bwb,bass}. In
Ref.\cite{esb} $S_{ct}$ was not factorised out of the  breakup amplitude but
since the results obtained both for the core momentum distribution and the
neutron breakup cross section are consistent with those of Ref.\cite{hans} and
with the results we present in this paper, the authors of \cite{esb} conclude
that the factorised form is a good approximation \cite{ge}. 

The effect of
$|S_{ct}|^2$ in the integral in Eq.(\ref{cross}) is to cut-off small values of
the core-target impact parameters. In Ref.\cite{27} we used a parametrized form
for $|S_{ct}|^2$ which corresponds to a smooth cut-off in  $d$ .
 The effect of the smooth cut-off depends on the parameter $\eta$
defined above. In the case of weak initial binding energy $\eta$ is very small
and the smooth cut-off correction is negligible as compared to a sharp cut-off.
Furthermore in a recent paper \cite{sl} it has been shown that the breakup
probability as a function of the core-target impact parameter, corresponding to
$P(\varepsilon_f,d) P_{el}(d)$ in our formalism, is peaked at a radius quite
larger than the sum of the projectile-target radii and it rises sharply from
zero to the maximum value (cf.Fig.2 of \cite{sl}). In \cite{sl} a DWBA approach
has been used, which takes into account properly the distortion of the relative
motion trajectory by the core-target optical potential.

The above discussion
justifies the use of a sharp cut-off in $d$.  Thus a  simple expression for
the cross section is obtained, which is valid  when $P_{el}$ is given  by the
strong absorption model such that $P_{el}=1$ if $d\geq R_s$ $P_{el}=0$ if $d<
R_s$ \cite{ab},
 \begin{equation} {d \sigma \over d \varepsilon_f}\approx 
C^2S{\pi R_s\over \eta}\left(1+ {1\over 2 \eta
R_s}\right) {d P (R_s) \over d\varepsilon_f}.\label
{tc}\end{equation}

The above equation is consistent to leading order in $1\over \eta$
with the formulae given in \cite{hans} for $l_i=0,1$.
  Eq.(\ref{pg}) is also similar to the breakup probability of
\cite{hans}  where instead of the Glauber elastic plus absorption
cross section factor the experimental neutron-target reaction cross
section was used. The breakup cross section is sensitive to the choice
of the strong absorption radius $R_s$ \cite{bass,DMB,sat}. It can be estimated
from 
 \begin {equation}
 R_s = 1.4 (A_t^{1/3} + A_{pc}^{1/3}) \ {\rm fm}. \label{strong}
 \end {equation}
where $A_t$ and $A_{pc}$ are the mass numbers of the target and projectile
core.

The momentum distribution is one factor in the breakup probabilities
expressions Eq.(\ref{dpde}) and (\ref{pg})  but there are
other factors which combine to modify its relation with the cross
section Eq.(\ref{tc}). One is the effect of the neutron-target
rescattering shown explicitly by the Glauber factors of
Eq.(\ref{pg}). Another is the $\eta$ dependence in the factor ${\pi
R_s\over \eta}\left(1+ {1\over 2 \eta R_s}\right) $ in Eq.(\ref{tc})
coming from the ion-ion scattering and giving a kind of modified
nucleus-nucleus geometrical cross section.

 The incident energy dependence in the total breakup probability is
contained mainly in the neutron-target phase-shift $\delta_{l_f}$, as
shown explicitly by the Glauber form $\chi({\bf b})\approx \tilde {V_2}({\bf b} ,
0)/(v\hbar) $.  This is partly because of the factor $1/v$ in the expression for
$\chi({\bf b})$ but also because the
  neutron-target optical potential $V_2$ is energy dependent.  However
  as we mentioned after Eq.(\ref{pg}) the
   energy distribution of the final neutron has its main origin in the 
  $ k_1$ dependence of  $\tilde {\psi}_1(d,k_1)$, the  momentum
  distribution of the neutron in the initial state.
   With the approximate form Eq.(\ref{psi}) it is contained mainly in
  the factor $\exp(-2\eta d)$.

  \section
  {\bf Application to $^{11}Be$ breakup}\label{three}

The situation
is especially striking for breakup of a halo nucleus like $^{11}Be$ 
where the binding energy
$\varepsilon_i=-0.5MeV$ of the neutron in the  initial nucleus is
very small. The factor  $\exp(-2\eta d)$ 
  has a sharp maximum when $\eta=\gamma_i=0.155 fm^{-1}$ or $k_1=0$ 
and $\varepsilon_f\approx mv^2/2$.  The small
value of $\gamma_i $ gives distributions  in $k_1$  which are sharper  than
those obtained for normal heavy  ions where the initial binding energy
is of the order of $-10MeV$ and
it makes the breakup cross section large. The values of 
$\eta$  as a function of the neutron final energy $\varepsilon_f$
are shown in Fig.(1a) for the case of $^{11}Be$ by the solid line.
Also shown is the same parameter for a $^{10}Be$ projectile (dot line) whose
 neutron separation energy is $\varepsilon_i=-6.8MeV$ and $l_i=1$.
There is a much stronger energy dependence and the minimum value of
$\eta$ is much smaller in the case of $^{11}Be$.  The corresponding
$k_1$-dependence of $\tilde {\psi}_1$, calculated from
Eqs.(\ref{psi}) is shown in Fig.(1b). As expected the momentum
distribution is sharper in $^{11}Be$ (solid line) than in $^{10}Be$
(dashed line).  It is also a characteristic of an s-state momentum
distribution to be narrower than for p-states, d-states or larger $l$-values.  

The initial state amplitude
$\tilde {\psi}_1(d,k_1)$ depends on the choice of $d=R_s$
which  in Fig.(1b) was taken to be  $6.2fm$ and which is appropriate
for the breakup reaction of $^{11}Be$   on a  $^{9}Be$ target. For a
$^{208}Pb$ target we take $R_s=11.5fm$. These are close to values given by
Eq.(\ref{strong}). The widths  at half maximum of
the resulting momentum distributions are   $\hbar \Delta  k_1=49MeV/c$
for a $^{9}Be$ target and $\hbar \Delta k_1=39.4MeV/c$ for a $^{208}Pb$
target. The difference is due to the change in the strong absorption
radius. The dependence of the parallel momentum distribution on $d$ or
equivalently on $R_s$ can be understood by noting that the parallel
momentum distribution Eq.(\ref{psi}) is  such that  the square of
its modulus represents the probability of finding a neutron in the
projectile with a certain value of its momentum parallel to the
relative motion velocity, when it is at a definite distance $d$
from the center of the initial nucleus.

The next step is to discuss whether information about the width of the
neutron momentum distribution in the initial nucleus can be obtained
from the breakup spectra. To be specific we focus on breakup reactions
with a $^{11}Be$ projectile on two targets. One possibility would be
to use the measured spectra of breakup neutrons. In fact most of the
present experiments use the energy loss spectra of the $^{10}Be$
fragments.  If elastic breakup is the only mechanism leading to
$^{10}Be$ fragments these two approaches would be equivalent. By
energy conservation, Eq.(\ref{ec}), the $^{10}Be$ energy loss spectra
would have the same shape as the neutron spectra of Fig.(2). On the
other hand the absorption component of Eqs.(\ref{dpde}) and (\ref{pg})
refers to neutrons which are detached from the projectile and are
either inelastically scattered or cause some reactions in the
target. They would have their energy degraded or might not been seen
at all, but the energy loss spectrum of the ejectile can still be
measured. In the figures the spectra are plotted as a function of 
$\varepsilon_f$, but they can be interpreted as energy loss spectra
using equation (\ref{ec}).

 In Fig.(2) we show the 
energy spectra  for the reaction  $^{9}Be( ^{11}Be,^{10}Be)^{9}Be+n$ at
$E_{inc}=10$, $41$ and $72A.MeV$ calculated according to
Eq.(\ref{dpde})  with an optical model S-matrix. The optical
potential used  is from \cite{pot1}. Technical details concerning
 this kind of calculations can be found in  Refs.\cite{bb1,bb3}.  For the $2s_{1/2}$ initial state in
$^{11}Be$ 
 the parameters appearing in Eqs.(\ref{psi}) and (\ref{tc}) are: the
spectroscopic
 factor $C^2S=0.77$, and the asymptotic normalization constant
$C_1=0.94 fm^{-1/2}$.
 The dot lines in Fig.(2) are the elastic breakup spectra
while the dashed lines   are the absorption spectra. 
The solid lines give their sum which correspond
 to the ejectile    inclusive energy spectra leading to $^{10}Be$
fragments.   The relationship between the momentum scale in Fig.(1b)
and the energy scale in Fig.(2) is contained in   the definition of
$k_1$ given after Eq.(\ref{B}). In the same way the neutron parallel
momentum distribution after breakup can be obtained from
Eq.(\ref{tc})  by a simple change of variable from $\varepsilon_f$ to
$k_1$. Thus the simple expectation is that the width
$\Delta\varepsilon_f$ of the neutron energy spectrum should be related
to the width $\hbar\Delta k_1$ of the halo neutron momentum
distribution by

\begin{equation} \hbar \Delta k_1=\Delta\varepsilon_f/
v.\label{ke}\end {equation}

However because of the extra factors in Eq.(\ref{tc}) there is a
modification in the above relation. To show the extent of the
modification Table I gives values of $\hbar \Delta k_1$ obtained from
the calculated energy spectra using the relation (\ref{ke}) for a 
$^{9}Be$ target and a $^{208}Pb$ target. The optical potentials used
were from \cite{pot1} and \cite{pot3} respectively.
In each column the first number refers to $^{208}Pb$
 and the second to $^{9}Be$ . The numbers in the last column should be
compared with the  widths of the momentum distribution Eq.(\ref{psi}),
which is $\hbar \Delta k_1=39.4MeV/c$ for the $^{208}Pb$ target,
because $R_s=11.5$ and $\hbar \Delta k_1=49MeV/c$ for the $^{9}Be$
target from Fig.(1b), where $R_s=6.2$.  In the case of Pb there is a
close correspondence between the width of the calculated energy
distribution and the width of the momentum distribution $\Delta
k_1$. On the other hand the values of $\Delta k_1$ extracted from the
energy distribution for the Be target are 20$\%$ smaller than the
$\Delta k_1$ from Fig.(1b). 

To show this effect more clearly we have plotted in Fig.(3a) the
differential cross section or energy spectra as a function of $k_1$,
at the same incident energies as Fig.(2).  Diamonds, dot line and
crosses are for $E_{inc}=10$ , $41$, $72A.MeV$ respectively. The full
line is the initial momentum distribution in $^{11}Be$ as in
Fig.(1b). All curves are normalized at the same peak value. Fig.(3a)
is for the Be target and one can see that the initial momentum
distribution is wider than the values obtained from the energy
spectra. For a Pb target instead, Fig.(3b) shows that the energy
spectra at high energy reproduce well the initial momentum
distribution width. The different scales in Figs.(3a) and (3b) reflect
the difference in $\tilde \psi_1(d,k_1)$ when $d$ changes.  One
possible reason for this behaviour is the fact that in the case of the
Pb target the halo wave function is probed at a larger distance as
compared to the Be case.
 Thus changing the target but keeping the incident energy fixed on can
get information on the momentum distribution at different
radii. 

Results which seem consistent with our discussion have been
obtained  by Kelley et al. \cite{ke} who measured the parallel
momentum distribution of $^{10}Be$ fragments on several targets. They found
$\Delta k_1=41.6\pm2.1MeV/c$ on a Be target, in agreement with our value in
Table I. Also their results show a $5\%$ change in the widths depending on the
target, again in agreement with the maximum variation shown by our
calculations. A similar target dependence was remarked in \cite{hjj} in
connection with $^{11}Li$ data. On the other hand the small $\Delta k_1$ values
obtained at low incident energies for the same targets are due to the fact that
because of the relation between $k_1$ and $\varepsilon_f$ the tails of the
momentum distributions in Fig.(1b) cannot be
 sampled by a low energy reaction. 

In the case of a heavy target the experimental spectra contain an
important contribution from Coulomb breakup which we discuss in
\cite{jm}. Here we anticipate that for the lead target the Coulomb
breakup calculated spectra give $\hbar\Delta k_1=43 MeV/c$ and $47
MeV/c$ at $E_{inc}=110 MeV$ and $790 MeV$ respectively.

 Widths calculated from the energy distributions obtained
with the eikonal model Eq.(\ref{pg}) are about 10$\%$ smaller than the ones
given  in Table I obtained with an optical model final wave function 
using the more accurate energy distribution in Eq.(\ref{dpde}). 
The agreement between the two calculations  improves  increasing the
incident energy and for heavy targets. However the absorption
is always underestimated by the eikonal approximation. This is
because the contribution from the very low partial waves which is
correctly described by the optical modes S-matrix in Eq.(\ref{dpde})
is instead underestimated by the eikonal S-matrix. Elastic breakup is
dominated by high partial waves which are well described by the eikonal
S-matrix. In this respect it can be useful to remark that while for a
light target partial waves up to  $l_f=7$ are enough to describe the
neutron rescattering on the target, in the case of the Pb target we
found that partial waves up to $l_f=25$ have to be taken into account. This is
possible and easy to do with the present method while it requires very
cumbersome calculations in other  methods like DWBA.

The relative amount of nuclear vs. Coulomb breakup in the
 total  breakup cross section, for heavy targets is currently a question of
debate for the experimental and theoretical implications\cite{dlv}. In the
case of heavy targets we are studying the important contribution to the
exclusive cross sections from Coulomb breakup in
\cite{jm}. The results could be used to disentangle the two
mechanisms.
Therefore  we give in Table II the calculated cross sections integrated over
energies and angles, for the breakup of 
$^{11}Be$ on $^{9}Be$, $^{48}Ti$ and $^{197}Au$ \cite{anne} at
$E_{inc}=41A.MeV$ and on a $^{208}Pb$ target \cite{nak} at
$E_{inc}=72A.MeV$ together with the experimental values for the inclusive and
exclusive cross sections. 
 The first row gives experimental and theoretical values for inclusive cross
sections. The theoretical values are the sum of the  elastic, Coulomb  and
inelastic contributions, given  in the third to fifth rows. The second row gives 
the exclusive values. In this case the theoretical estimates are 
 the sum of the Coulomb and nuclear elastic cross sections. The Coulomb breakup
calculations \cite{jm} were done in first order perturbation theory similarly to
\cite{anne}. Our results agree well with experimental values and with other
theoretical estimates \cite{anne,bj} and show that for heavy targets the nuclear
breakup, although  smaller than the Coulomb breakup, cannot be considered
negligible. The cross section for the lead target are slightly smaller than
those for the gold target because the incident energy was higher.

Our results for the reaction 
$^{9}Be(^{11}Be,^{10}Be)^{9}Be+n$ were obtained with $R_s=6.2fm$. Using
$R_s=5.9fm$ gives $\sigma_{breakup}=0.41 b$ while using $R_s=6.5fm$ gives
$\sigma_{breakup}=0.36 b$.
 For comparison, the results for the neutron breakup from the
$^{10}Be$ core at $R_s=6.2fm$ are: $\sigma^{el}=0.025b$ and
$\sigma^{inel}=0.037b$, where we assumed a spectroscopic factor $C^2S=4$ for
the $1p_{3/2}$ state in $^{10}Be$. Since the experimental spectroscopic factor
is in fact 2.1 and the rest of the occupation probability is spread over three
excited bound states  in $^{10}Be$, the above value can be considered as a
higher limit estimate for the core breakup contribution at $E_{inc}=41A.MeV$.
The elastic plus inelastic cross sections obtained using the eikonal formula 
(\ref{pg})  are smaller by about 25$\%$.  This is mainly due to the absorption
term for which the eikonal approximation is not very accurate. Here and in the
following we  refer to the cross section coming from the absorptive term of
Eq.(\ref{dpde}) as inelastic.

 At this point it is worth mentioning that some
authors\cite{esb,hans} refer to the absorptive term as to stripping, while  
others \cite{ot} have used the term inelastic breakup to include also
processes in which the core nucleus interacts inelastically with the target,
besides those discussed here which are due to the neutron-target interaction.
However as it has been discussed in \cite{ot} these processes are expected to
give rise to a small core survival probability and therefore they have often
been neglected. Since it is always the $(A_p-1)$ nucleus which is detected, its
possible excitations must be restricted to those below particle threshold. We
have discussed and included such processes in our calculations for breakup of
normal nuclei \cite{bb3,tiina} and the same can be done for halo projectiles. The
values of  cross sections for the core breakup given above give an estimate of
such processes. Some experimental evidence for such reactions has recently been
discussed by Hansen\cite{varhan}.

In view of future experiments  in the incident energy range
$3-10 A.MeV$ it is useful to know the energy dependence of
the nuclear elastic  and inelastic  total breakup cross section.   In  Fig.(4a) we show the total elastic (cross) and
inelastic (star) cross sections and the sum (diamond) of them for a
$^{9}Be$ target.  They have been obtained from the integration of the
energy spectra from Eqs.(\ref{dpde}) and (\ref{tc}) calculated at
different incident energy.  The very interesting point is the relative
behaviour at low energy of the two components of the breakup and also the
dependence on the target. For this reason we show also in Figs.(4b) and (4c) the
cross section values calculated for  a $^{28}Si$\cite{flo}  and a 
$^{208}Pb$\cite{nak} target. The
neutron optical potentials used are from \cite{pot1,pot2,pot3} respectively, for
the  three targets. Comparing the three cases we notice that at high
incident energies the elastic  and inelastic  breakup tend to give the
same cross section. This is the geometric limit of the nuclear cross
section. At low energies in the case of the light target $^{9}Be$ the
elastic breakup is larger than the inelastic one and this pattern is
maintained increasing the incident energy. This is because the
neutron-light-target optical potentials have only a surface
component of the imaginary part, due to the fact that these nuclei can
have surface collective excitations but do not have enough density of
levels to allow the absorption into compound nucleus. On the other
hand in the case of $^{208}Pb$ the absorption is larger at all
energies  because there is a large probability of surface excitations
leading to neutron inelastic breakup and also to  the fact that the
neutron-Pb optical potential has  a volume term of almost constant
magnitude at high energies. Our results are consistent with those of
\cite{esb,hans} in the case of the Be target.


\section
{\bf Conclusions}\label{con}

To conclude, neutron breakup from the projectile in a
heavy-ion reaction 
 has been discussed in this paper in the framework 
of the theory of transfer  to the continuum reactions which uses an
optical model final state wave function for the neutron and the
relation with the eikonal model has been clarified.
The latter is a good guideline to understand the mechanism of halo breakup and
it is quite accurate to describe the halo neutron  breakup at high
incident energy and on a heavy target. In other situations one should 
use Eq.(\ref{dpde}) with an optical model S-matrix. 
  Both methods can be used to
calculate energy spectra but the  angular distributions of the breakup
neutron can be obtained only from the eikonal model   \cite {ab}
which contains the
 explicit dependence on the neutron final
momentum components, furthermore this method can
be easily extended  to relativistic energies, as it has been done in
\cite{esb}.

The models have been applied to the breakup of $^{11}Be$ on several target
nuclei. The neutron energy distribution has a peak centered at the incident
energy per nucleon of the $^{11}Be$ projectile with a width related to the
width of the momentum distribution of the halo neutron. To a first
approximation there is a very simple relation between these two widths, but
sample calculations show that, due to details of the
reaction mechanism, there can be differences between the initial momentum
distribution and the measured one which can reduce its original widths  up to
20$\%$ for a light target nucleus. 

Integrated nuclear elastic and inelastic 
breakup cross sections have been calculated and compared to experimental values
and  to Coulomb breakup cross sections. In the case of heavy targets it is very
important to estimate the two contributions separately. Our results agree well
with experimental values and show that for heavy targets the nuclear breakup,
although  smaller than the Coulomb breakup, cannot be considered negligible.
Finaly the  incident energy dependence of the nuclear elastic, inelastic and
total cross section has been studied.

We are grateful to P.G.Hansen for comments on an early version of this
paper and to G.F. Bertsch for discussions. One of us (A.B.) wishes to
thank W.Tornow for discussions and for information on
 neutron-light-nucleus optical potentials and H.Horiuchi for the 
hospitality in Kyoto during the preparation of this paper.

\newpage

\small{{{\bf Table I:}
Parallel momentum distribution widths for Be and Pb targets.  The
value of $\hbar v$ is related to the incident energy but takes into
account the effect of the nucleus-nucleus Coulomb barrier at $R_s$}}
\begin {center}
 \begin{tabular}[bht]{|c|c|c|c|}  \hline
 $E_{inc}$&  $\Delta \varepsilon_f$&$\hbar v$&$\hbar \Delta k_1$\\
(MeV)&(MeV)&(MeV  $\cdot$ fm)&(MeV/c) \\  \hline\hline
$110$&4.4~~5.0&22.45~~27.71&38.6~~35.5\\
$300$&8.8~~9.0&44.03~~46.92&39.3~~37.8\\
$790$&15.0~~15.5&75.1~~76.82&39.3~~39.8\\ \hline\hline
Target&Pb~~~Be&Pb~~~~Be&Pb~~~Be\\
\hline\end{tabular}\end{center}

\vskip 0.5in
\small{{{\bf Table II:}
Cross sections for Be, Ti, Au\cite {anne} and Pb\cite{nak} targets.  The first
 row gives experimental and theoretical values for inclusive cross sections. The
theoretical values are the sum of the  elastic, Coulomb  and inelastic
 contributions, given  in the third to fifth rows. The second row gives 
the exclusive values. In this case the theoretical estimates are 
 the sum of the Coulomb and nuclear elastic cross sections.}

\begin{tabular}[bht]{|c|c|c|c|c|}  \hline
 $\sigma(b)$& Be&Ti&Au&Pb\\
  \hline\hline
$(^{11}Be,^{10}Be)$&$0.29\pm 0.04
$~~0.39b&$0.65\pm 09$~~0.70&$2.45\pm0.20$~~2.73&----~~2.39\\
$(^{11}Be,^{10}Be+n)$&$0.24\pm 0.05$~~0.23&$0.55\pm 0.11~~0.47$&
$2.5\pm 0.5~~2.37$&$1.8\pm 0.4~~2.09$\\
$el$&0.22&0.28&0.35&0.3\\ 
$Coul$&0.0074&0.19&2.02&1.79 \\
$inel$&0.16&0.234&0.39&0.3\\ \hline\hline
\end{tabular}

\newpage

{\bf Figure Captions}
\begin{description}
\item{\bf Fig.1}.(a) Values of the parameter $\eta$ from  for 
 $^{11}Be$ solid line, and $^{10}Be$ dot line.
(b) Initial momentum distribution for the $2s_{1/2}$ state in
$^{11}Be$, solid line, and $1p_{1/2}$-state in $^{10}Be$, dot line.

 \item{\bf Fig.2}. Final energy spectra from Eq.(\ref{tc})
for $^{9}Be(^{11}Be,^{10}Be)^{9}Be+n$ at $E_{inc}=10$ ,  $41$,
$72A.MeV$. Dot lines are for elastic breakup, dashed lines for
inelastic breakup and solid lines are their sum corresponding to
the inclusive $^{10}Be$ spectrum via Eq.(\ref{ec}).

\item{\bf Fig.3} (a). Final  energy spectra as a function of
$k_1$, at the same incident energies as Fig.(2). Diamonds, dot line
and crosses are for $E_{inc}=10$, $41$, $72A.MeV$ respectively. Solid
line is the initial momentum distribution as in Fig.(1b). All curves
are normalized at the some peak value. (b) The same  as Fig.(3a) for
$^{208}Pb(^{11}Be,^{10}Be)^{208}Pb+n$.

\item{\bf Fig.4} Incident energy dependence of the breakup cross sections
on $^{9}Be$ (4a),$^{28}Si$ (4b)  and $^{208}Pb$ (4c).
Diamonds joined by full lines give the total breakup cross sections.
 Crosses with dot lines are for the elastic
breakup  while stars with dotdashed lines are the inelastic breakup cross sections.

 \end{description}

 
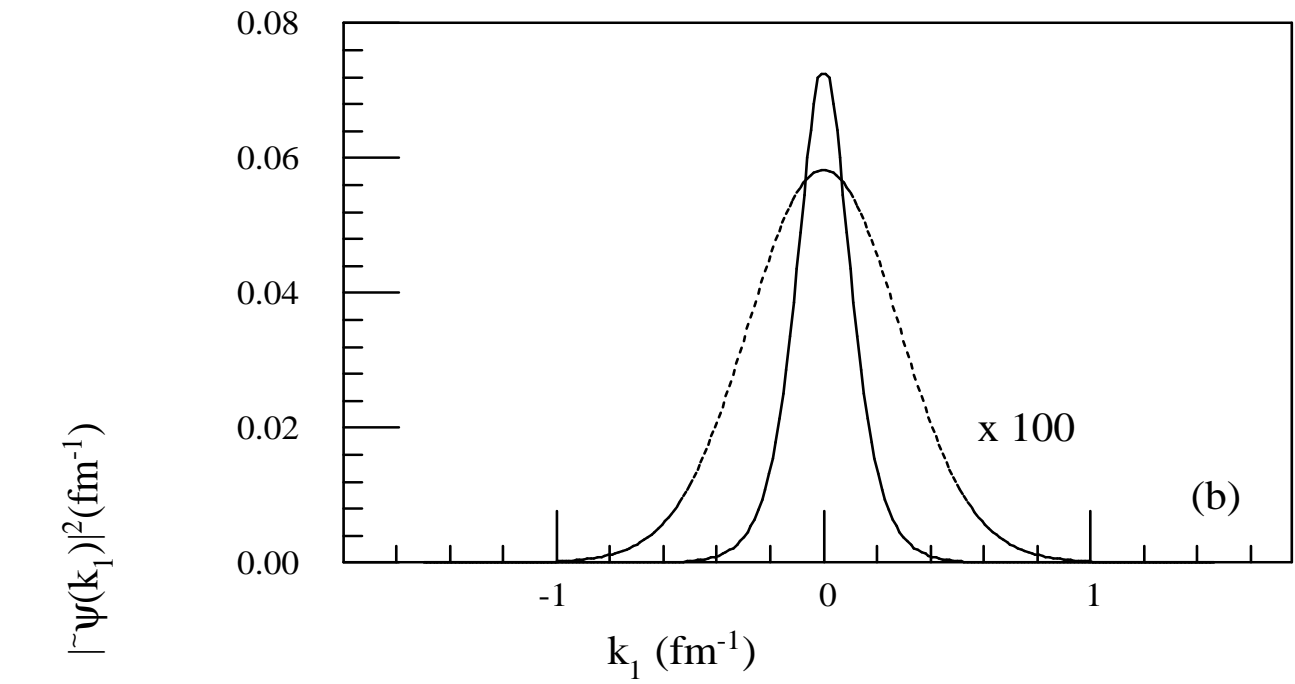
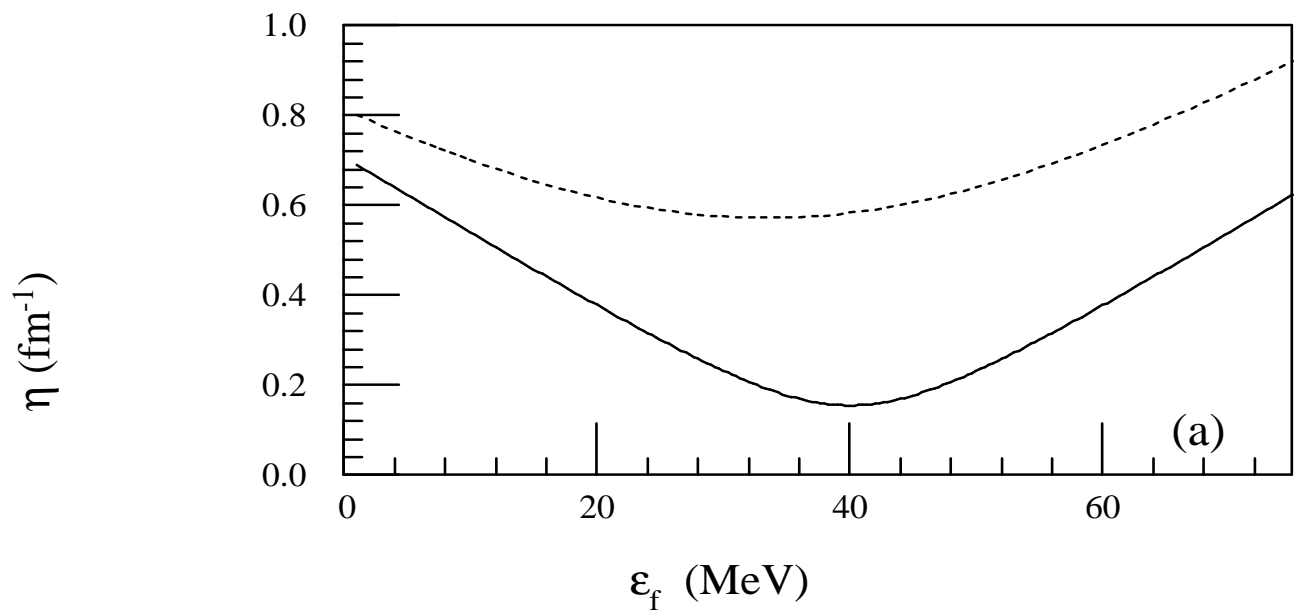
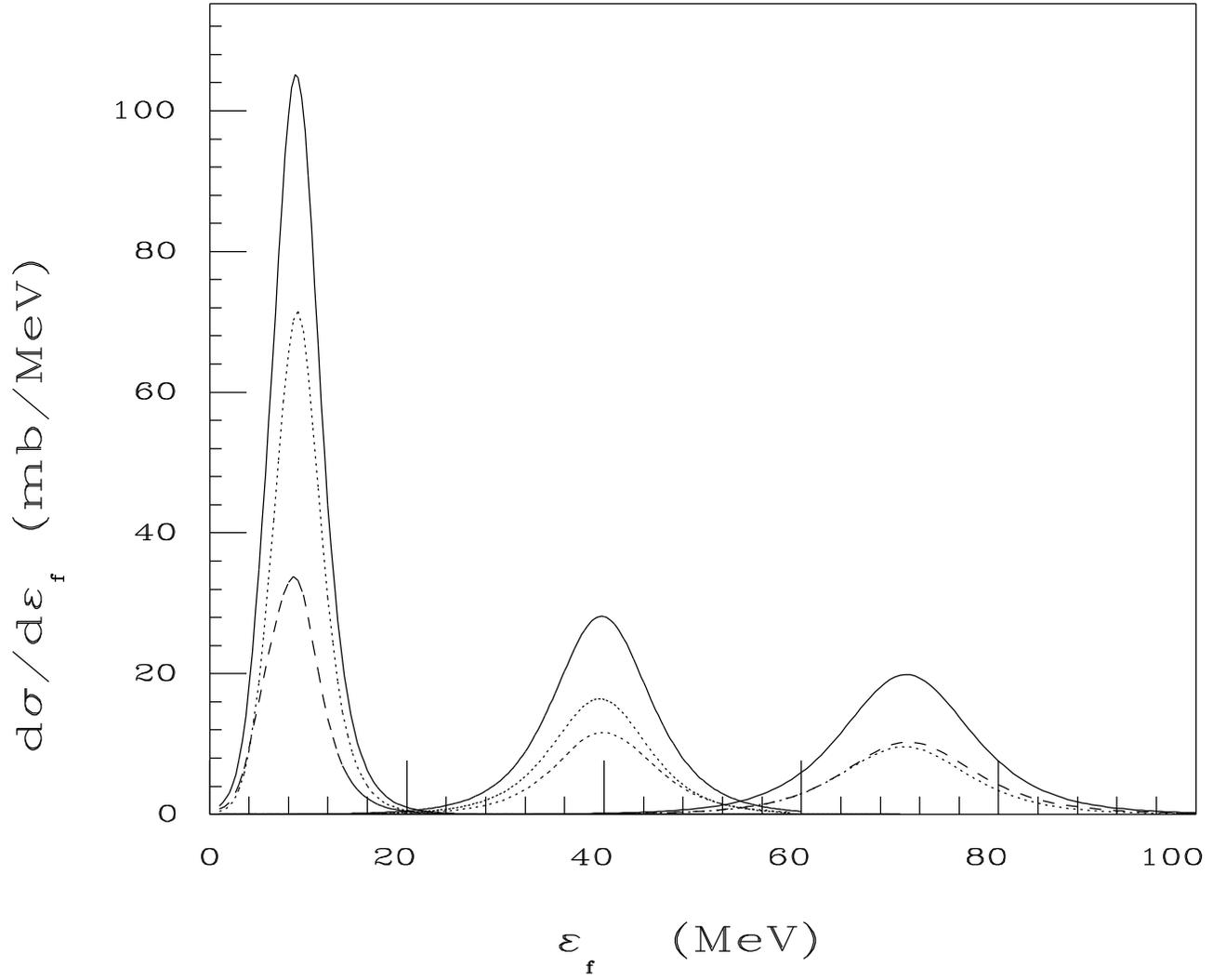
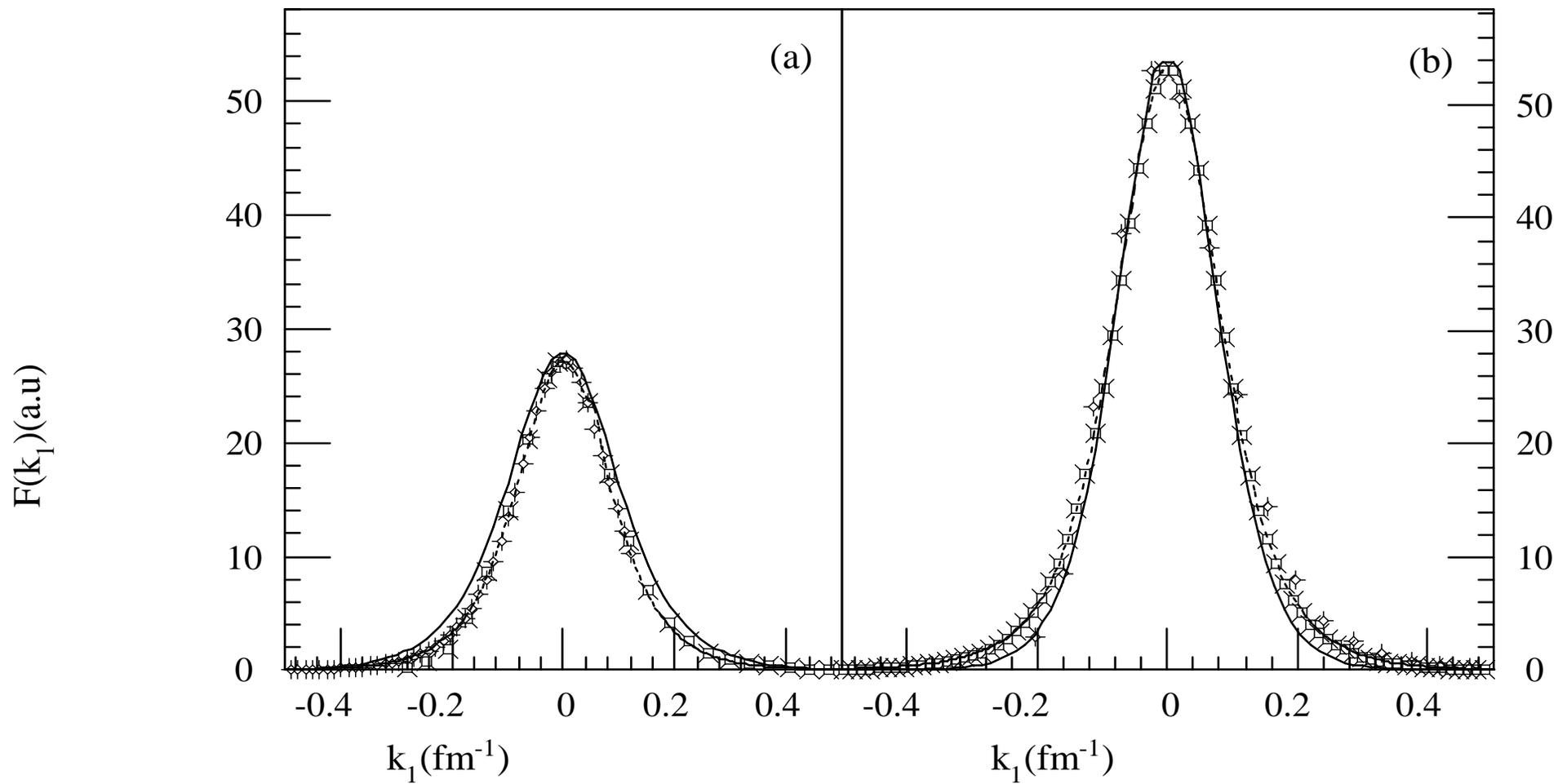
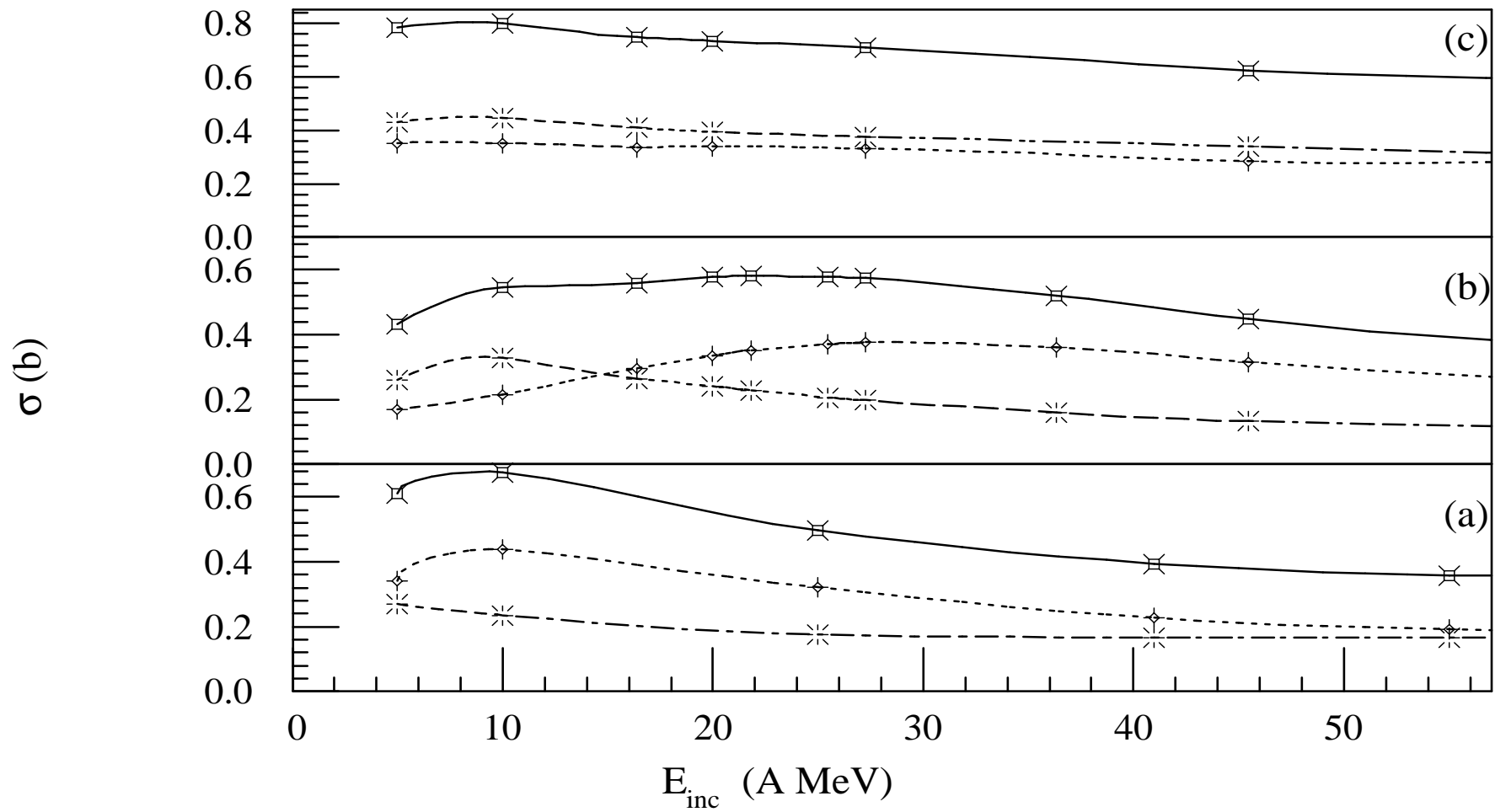
\end{document}